\RequirePackage{amsmath}
\RequirePackage{fix-cm}
\documentclass[smallextended]{svjour3}
\smartqed
\usepackage{amsfonts}
\usepackage{cite}
\usepackage[colorlinks=true]{hyperref}
\newtheorem{conj}[theorem]{Conjecture}

\journalname{Letters in Mathematical Physics}
\begin{document}

\title{The Wehrl entropy has Gaussian optimizers\thanks{I acknowledge financial support from the European Research Council (ERC Grant Agreement no 337603), the Danish Council for Independent Research (Sapere Aude) and VILLUM FONDEN via the QMATH Centre of Excellence (Grant No. 10059).}}

\author{Giacomo De Palma}

\institute{Giacomo De Palma \at QMATH, Department of Mathematical Sciences, University of Copenhagen, Universitetsparken 5, 2100 Copenhagen, Denmark\\ Tel.: +45 35 33 68 04\\ \email{giacomo.depalma@math.ku.dk}}

\date{}
\maketitle

\begin{abstract}
We determine the minimum Wehrl entropy among the quantum states with a given von Neumann entropy, and prove that it is achieved by thermal Gaussian states.
This result determines the relation between the von Neumann and the Wehrl entropies.
The key idea is proving that the quantum-classical channel that associates to a quantum state its Husimi $Q$ representation is asymptotically equivalent to the Gaussian quantum-limited amplifier with infinite amplification parameter.
This equivalence also permits to determine the $p\to q$ norms of the aforementioned quantum-classical channel in the two particular cases of one mode and $p=q$, and prove that they are achieved by thermal Gaussian states.
The same equivalence permits to prove that the Husimi $Q$ representation of a one-mode passive state (i.e. a state diagonal in the Fock basis with eigenvalues decreasing as the energy increases) majorizes the Husimi $Q$ representation of any other one-mode state with the same spectrum, i.e. it maximizes any convex functional.
\end{abstract}

\section{Introduction}
The Husimi $Q$ representation \cite{husimi1940some} is a probability distribution in phase space that describes a quantum state of a Gaussian quantum system, such as an harmonic oscillator or a mode of electromagnetic radiation \cite{leonhardt1997measuring,barnett2002methods}.
It coincides with the probability distribution of the outcomes of a heterodyne measurement \cite{schleich2015quantum} performed on the state.
This measurement is fundamental in the field of quantum optics.
It is used for quantum tomography \cite{carmichael2013statistical}, and it lies at the basis of an easily realizable quantum key distribution scheme \cite{weedbrook2012gaussian,weedbrook2004quantum}.
The Husimi $Q$ representation is also used to study quantum effects in superconductors \cite{callaway1990remarkable}.

The Wehrl entropy \cite{wehrl1979relation,wehrl1978general} of a quantum state is the Shannon differential entropy \cite{cover2006elements} of its Husimi $Q$ representation.
It is considered as the classical entropy of the state, and it coincides with the Shannon differential entropy of the outcome of a heterodyne measurement performed on the state.

While the von Neumann entropy of any pure state is zero, this is not the case for the Wehrl entropy.
E. Lieb proved \cite{lieb1978proof,carlen1991some} that the Wehrl entropy is minimized by the Glauber coherent states \cite{schrodinger1926stetige,bargmann1961hilbert,klauder1960action,glauber1963coherent,klauder2006fundamentals}, with a proof based on some difficult theorems in Fourier analysis.
This result has then been generalized to symmetric $SU(N)$ coherent states \cite{lieb2015proof,lieb2014proof}.
It has also been proven \cite{lieb2014proof,giovannetti2015majorization,holevo2015gaussian} that the Husimi $Q$ representation of coherent states majorizes the Husimi $Q$ representation of any other quantum state, i.e. it maximizes any convex functional.

Let us now suppose to fix the von Neumann entropy of the quantum state.
What is its minimum possible Wehrl entropy?
We determine it and prove that it is achieved by the thermal Gaussian state with the given von Neumann entropy (\autoref{thm:entropy}).
This result determines the relation between the von Neumann and the Wehrl entropies.
The key idea is proving that the quantum-classical channel that associates to a quantum state its Husimi $Q$ representation is asymptotically equivalent to the Gaussian quantum-limited amplifier with infinite amplification parameter (\autoref{thm:HA}).
We can then link this equivalence to the recent results on quantum Gaussian channels \cite{de2015passive,de2016gaussian,de2016pq,de2016gaussiannew,frank2017norms,holevo2017quantum}, noncommutative generalizations of the theorems in Fourier analysis used in Lieb's original proof.
This link also permits to determine the $p\to q$ norms of the aforementioned quantum-classical channel in the two particular cases of one mode and $p=q$, and prove that they are finite for $1\le p\le q$, infinite for $p>q\ge 1$, and in any case achieved by thermal Gaussian states (\autoref{thm:pq}).
Moreover, the same link implies that the Husimi $Q$ representation of a one-mode passive state (i.e. a state diagonal in the Fock basis with the eigenvalues decreasing as the energy increases) majorizes the Husimi $Q$ representation of any other one-mode state with the same spectrum, i.e. it maximizes any convex functional (\autoref{thm:maj}).

The paper is structured as follows.
\autoref{sec:GQS} introduces Gaussian quantum systems and Gaussian quantum states, and \autoref{sec:husimi} introduces the Husimi $Q$ representation.
\autoref{sec:A} defines the Gaussian quantum-limited amplifier and presents the recent results on quantum Gaussian channels needed for the proofs.
\autoref{sec:main} presents our results, that are proved in \autoref{sec:HA}, \autoref{sec:maj}, \autoref{sec:pq} and \autoref{sec:entropy}.
\autoref{sec:concl} draws the conclusions.
\autoref{app:E} defines the Gaussian quantum-limited attenuator, that is needed for some of the proofs.
\autoref{app:aux} contains some auxiliary theorems and lemmas.

\section{Gaussian quantum systems}\label{sec:GQS}
We consider the Hilbert space of $M$ harmonic oscillators, or $M$ modes of the electromagnetic radiation, i.e. the irreducible representation of the canonical commutation relations
\begin{equation}
\left[\hat{a}_i,\;\hat{a}_j^\dag\right]=\delta_{ij}\;\hat{\mathbb{I}}\;,\qquad i,\,j=1,\ldots,\,M\;.
\end{equation}
The operators $\hat{a}_1^\dag\hat{a}_1,\ldots,\,\hat{a}_M^\dag\hat{a}_M$ have integer spectrum and commute.
Their joint eigenbasis is the Fock basis $\left\{|n_1\ldots n_M\rangle\right\}_{n_1,\ldots,\,n_M\in\mathbb{N}}$.
The Hamiltonian
\begin{equation}
\hat{N}=\sum_{i=1}^M\hat{a}_i^\dag\hat{a}_i
\end{equation}
counts the number of excitations, or photons.

One-mode thermal Gaussian states have density matrix
\begin{equation}\label{eq:omegaz}
\hat{\omega}_z=\sum_{n=0}^\infty \left(1-z\right)z^n|n\rangle\langle n|\;,\qquad 0\le z<1\;.
\end{equation}
Their average energy is
\begin{equation}\label{eq:Ez}
E = \mathrm{Tr}\left[\hat{N}\;\hat{\omega}_z\right] = \frac{z}{1-z}\;,
\end{equation}
and their von Neumann entropy is
\begin{equation}\label{eq:defg}
S = -\mathrm{Tr}\left[\hat{\omega}_z\ln\hat{\omega}_z\right] = \left(E+1\right)\ln\left(E+1\right)-E\ln E:=g(E)\;.
\end{equation}

\section{The Husimi \emph{Q} representation and Wehrl entropy}\label{sec:husimi}
The classical phase space associated to a $M$-mode Gaussian quantum system is $\mathbb{C}^M$, and for any $\mathbf{z}\in\mathbb{C}^M$ we define the coherent state
\begin{equation}
|\mathbf{z}\rangle = e^{-\frac{|\mathbf{z}|^2}{2}}\sum_{n_1,\ldots,\,n_M\in\mathbb{N}}\frac{z_1^{n_1}\ldots z_M^{n_M}}{\sqrt{n_1!\ldots n_M!}}\;|n_1\ldots n_M\rangle\;.
\end{equation}
Coherent states are not orthogonal:
\begin{equation}\label{eq:cohov}
\langle\mathbf{z}|\mathbf{w}\rangle = e^{\mathbf{z}^\dag\mathbf{w}-\frac{|\mathbf{z}|^2+|\mathbf{w}|^2}{2}}\qquad\forall\;\mathbf{z},\,\mathbf{w}\in\mathbb{C}^M\;,
\end{equation}
but they are complete and satisfy the resolution of the identity \cite{holevo2015gaussian}
\begin{equation}\label{eq:complz}
\int_{\mathbb{C}^M}|\mathbf{z}\rangle\langle \mathbf{z}|\;\frac{\mathrm{d}^{2M}z}{\pi^M} = \hat{\mathbb{I}}\;,
\end{equation}
where the integral converges in the weak topology.
The POVM associated with the resolution of the identity \eqref{eq:complz} is called heterodyne measurement \cite{schleich2015quantum}.
\begin{definition}[Husimi $Q$ representation]
The Husimi $Q$ representation of a quantum state $\hat{\rho}$ is the probability distribution on phase space of the outcome of an heterodyne measurement performed on $\hat{\rho}$, with density
\begin{equation}
Q\left(\hat{\rho}\right)(\mathbf{z}) := \langle\mathbf{z}|\hat{\rho}|\mathbf{z}\rangle\;,\qquad \mathbf{z}\in\mathbb{C}^M\;,\qquad\int_{\mathbb{C}^M}Q\left(\hat{\rho}\right)(\mathbf{z})\;\frac{\mathrm{d}^{2M}z}{\pi^M} = 1\;.
\end{equation}
\end{definition}
\begin{definition}[Wehrl entropy]
The Wehrl entropy of a quantum state $\hat{\rho}$ is the Shannon differential entropy of its Husimi $Q$ representation
\begin{equation}
W\left(\hat{\rho}\right) := -\int_{\mathbb{C}^M} Q(\hat{\rho})(\mathbf{z})\;\ln Q(\hat{\rho})(\mathbf{z})\;\frac{\mathrm{d}^{2M}z}{\pi^M}\;.
\end{equation}
\end{definition}

\section{The Gaussian quantum-limited amplifier}\label{sec:A}
The $M$-mode Gaussian quantum-limited amplifier $\mathcal{A}_{\kappa}^{\otimes M}$ with amplification parameter $\kappa\ge1$ performs a two-mode squeezing on the input state $\hat{\rho}$ and the vacuum state of a $M$-mode ancillary Gaussian system $B$ with ladder operators $\hat{b}_1,\ldots,\,\hat{b}_M$:
\begin{equation}
\mathcal{A}_{\kappa}^{\otimes M}\left(\hat{\rho}\right)= \mathrm{Tr}_B\left[\hat{U}_\kappa\left(\hat{\rho}\otimes|\mathbf{0}\rangle\langle\mathbf{0}|\right)\hat{U}_\kappa^\dag\right]\;.
\end{equation}
The squeezing unitary operator
\begin{equation}\label{eq:defUk}
\hat{U}_\kappa=\exp\left(\mathrm{arccosh}\sqrt{\kappa}\sum_{i=1}^M\left(\hat{a}_i^\dag\hat{b}_i^\dag-\hat{a}_i\,\hat{b}_i\right)\right)
\end{equation}
acts on the ladder operators as
\begin{align}
\hat{U}_\kappa^\dag\;\hat{a}_i\;\hat{U}_\kappa &= \sqrt{\kappa}\;\hat{a}_i + \sqrt{\kappa-1}\;\hat{b}_i^\dag\;,\\
\hat{U}_\kappa^\dag\;\hat{b}_i\;\hat{U}_\kappa &= \sqrt{\kappa-1}\;\hat{a}_i^\dag + \sqrt{\kappa}\;\hat{b}_i\;,\qquad i=1,\ldots,\,M\;.
\end{align}
The Gaussian quantum-limited amplifier preserves the set of thermal Gaussian states, i.e. for any $0\le z<1$
\begin{equation}\label{eq:Az}
\mathcal{A}_\kappa\left(\hat{\omega}_z\right) = \hat{\omega}_{z'}\;,\qquad z'=1-\frac{1-z}{\kappa}\;.
\end{equation}

We now recall the latest results on quantum Gaussian channels, on which the proofs of \autoref{thm:maj}, \autoref{thm:pq} and \autoref{thm:entropy} are based.
The proof of \autoref{thm:maj} is based on
\begin{definition}[Passive rearrangement \cite{de2015passive}]
The passive rearrangement of the quantum state
\begin{equation}
\hat{\rho} = \sum_{n=0}^\infty p_n\;|\psi_n\rangle\langle\psi_n|\;,\quad p_0\ge p_1\ge\ldots\ge0\;,\quad \langle\psi_m|\psi_n\rangle=\delta_{mn}
\end{equation}
of a one-mode Gaussian quantum system is the state with the same spectrum with minimum average energy, i.e.
\begin{equation}
\hat{\rho}^\downarrow:=\sum_{n=0}^\infty p_n\;|n\rangle\langle n|\;.
\end{equation}
We say that $\hat{\rho}$ is \emph{passive} if $\hat{\rho} = \hat{\rho}^\downarrow$, i.e. if $\hat{\rho}$ is diagonal in the Fock basis with eigenvalues decreasing as the energy increases.
\end{definition}
\begin{theorem}[\cite{de2015passive}]\label{thm:majaux}
For $M=1$, the output generated by a passive input state majorizes the output generated by any other input state with the same spectrum, i.e. for any quantum state $\hat{\rho}$ and any convex function $f:[0,1]\to\mathbb{R}$ with $f(0)=0$
\begin{equation}
\mathrm{Tr}\;f\left(\mathcal{A}_\kappa\left(\hat{\rho}\right)\right) \le \mathrm{Tr}\;f\left(\mathcal{A}_\kappa\left(\hat{\rho}^\downarrow\right)\right)\;.
\end{equation}
\begin{remark}
\autoref{thm:majaux} does not hold for $M>1$ \cite{de2016passive}.
\end{remark}
\end{theorem}

The proof of \autoref{thm:pq} is based on the following conjecture, that has been proven in some particular cases \cite{giovannetti2015majorization,holevo2015gaussian,de2016pq,frank2017norms,holevo2017quantum}.
\begin{definition}[Schatten norm \cite{schatten1960norm,holevo2006multiplicativity}]
For any $p\ge1$ the Schatten $p$ norm of the positive semidefinite operator $\hat{A}$ is
\begin{equation}
\left\|\hat{A}\right\|_p := \left(\mathrm{Tr}\;\hat{A}^p\right)^\frac{1}{p}\;.
\end{equation}
\end{definition}
\begin{conj}\label{conj:pq}
For any $\kappa\ge1$ and any $p,\,q\ge1$ the $p\to q$ norm of $\mathcal{A}_\kappa^{\otimes M}$ is achieved by thermal Gaussian states, i.e. for any quantum state $\hat{\rho}$
\begin{equation}\label{eq:Apq}
\frac{\left\|\mathcal{A}_\kappa^{\otimes M}\left(\hat{\rho}\right)\right\|_q}{\left\|\hat{\rho}\right\|_p} \le \left(\sup_{0\le z<1}\frac{\left\|\mathcal{A}_\kappa\left(\hat{\omega}_z\right)\right\|_q}{\left\|\hat{\omega}_z\right\|_p}\right)^M\;.
\end{equation}
\begin{remark}\label{rem:pq}
Conjecture \ref{conj:pq} has been proven in Ref. \cite{de2016pq} for $M=1$, in Refs. \cite{frank2017norms,holevo2017quantum} for $p=q$ and any $M$, and in Refs. \cite{giovannetti2015majorization,holevo2015gaussian} for $p=1$ and any $M$.
For $p=1$, the supremum in \eqref{eq:Apq} is achieved by the vacuum state, i.e. in $z=0$.
For $p=q$, the supremum in \eqref{eq:Apq} is asymptotically achieved by the sequence of thermal Gaussian states with infinite temperature, i.e. for $z\to1$.
\end{remark}
\end{conj}
The proof of \autoref{thm:entropy} is based on the following fundamental result.
\begin{theorem}[\cite{de2016gaussiannew}]\label{thm:epni}
Thermal Gaussian states minimize the output von Neumann entropy of the one-mode Gaussian quantum-limited amplifier among all the input states with a given entropy, i.e. for any quantum state $\hat{\rho}$
\begin{equation}
S\left(\mathcal{A}_\kappa\left(\hat{\rho}\right)\right) \ge g\left(\kappa\;g^{-1}\left(S\left(\hat{\rho}\right)\right)+\kappa-1\right)\;,
\end{equation}
with $g$ as in \eqref{eq:defg}, and where $S(\hat{\rho})=-\mathrm{Tr}\left[\hat{\rho}\ln\hat{\rho}\right]$.
\end{theorem}

\section{Main results}\label{sec:main}
We start with the asymptotic equivalence between the Husimi $Q$ representation and the Gaussian quantum-limited amplifier, the key idea of the proofs of the other results.
\begin{theorem}[Husimi-amplifier equivalence]\label{thm:HA}
The quantum-classical channel that associates to a quantum state its Husimi $Q$ representation is asymptotically equivalent to the Gaussian quantum-limited amplifier with infinite amplification parameter, i.e. for any quantum state $\hat{\rho}$ and any real convex function $f\in C^1([0,1])$ with $f(0)=0$
\begin{equation}\label{eq:HA}
\int_{\mathbb{C}^M}f\left(\langle\mathbf{z}|\hat{\rho}|\mathbf{z}\rangle\right)\frac{\mathrm{d}^{2M}z}{\pi^M} = \lim_{\kappa\to\infty}\frac{\mathrm{Tr}\;f\left(\kappa^M\mathcal{A}_\kappa^{\otimes M}\left(\hat{\rho}\right)\right)}{\kappa^M}\;.
\end{equation}
\begin{proof}
See \autoref{sec:HA}.
\end{proof}
\end{theorem}
\begin{remark}
Since for any quantum state $\hat{\rho}$
\begin{equation}
0\le\langle\mathbf{z}|\hat{\rho}|\mathbf{z}\rangle\le1\;,
\end{equation}
and from \autoref{lem:Akbound}
\begin{equation}
0\le \kappa^M\mathcal{A}_\kappa^{\otimes M}\left(\hat{\rho}\right) \le \hat{\mathbb{I}}\;,
\end{equation}
the function $f$ in \autoref{thm:HA} needs to be defined in $[0,1]$ only.
\end{remark}
\begin{theorem}[majorization]\label{thm:maj}
For $M=1$, the Husimi $Q$ representation of a passive state majorizes the Husimi $Q$ representation of any other state with the same spectrum, i.e. for any quantum state $\hat{\rho}$ and any real convex function $f\in C^1([0,1])$ with $f(0)=0$
\begin{equation}
\int_{\mathbb{C}}f\left(\langle z|\hat{\rho}|z\rangle\right)\frac{\mathrm{d}^{2}z}{\pi} \le \int_{\mathbb{C}}f\left(\langle z|\hat{\rho}^\downarrow|z\rangle\right)\frac{\mathrm{d}^{2}z}{\pi}\;.
\end{equation}
\begin{proof}
See \autoref{sec:maj}.
\end{proof}
\begin{remark}
We state \autoref{thm:maj} for $M=1$ only since its proof relies on \autoref{thm:majaux}, that does not hold for $M>1$.
\end{remark}
\end{theorem}
\begin{theorem}[$p\to q$ norms]\label{thm:pq}
Assuming Conjecture \ref{conj:pq}, for any $1\le p\le q$ the $p\to q$ norm of the quantum-classical channel that associates to a quantum state its Husimi $Q$ representation is achieved by thermal Gaussian states, i.e. for any $M$-mode quantum state $\hat{\rho}$
\begin{equation}\label{eq:pq}
\frac{\left\|Q\left(\hat{\rho}\right)\right\|_q}{\left\|\hat{\rho}\right\|_p} \le \left(\sup_{0\le z<1} \frac{\left\|Q\left(\hat{\omega}_z\right)\right\|_q}{\left\|\hat{\omega}_z\right\|_p}\right)^M = \left(\sup_{0\le z<1}\frac{\left(1-z^p\right)^\frac{1}{p}}{q^\frac{1}{q}\left(1-z\right)^\frac{1}{q}}\right)^M\;,
\end{equation}
where
\begin{equation}
\left\|Q\left(\hat{\rho}\right)\right\|_q = \left(\int_{\mathbb{C}^M}{Q(\hat{\rho})(\mathbf{z})}^q\;\frac{\mathrm{d}^{2M}z}{\pi^M}\right)^\frac{1}{q}
\end{equation}
is the norm of $Q(\hat{\rho})$ in $L^q\left(\mathbb{C}^M\right)$.
The supremum in \eqref{eq:pq} and hence the $p\to q$ norm are finite if $1\le p\le q$, and infinite if $p>q\ge 1$.
\begin{proof}
See \autoref{sec:pq}.
\end{proof}
\begin{remark}
We recall that Conjecture \ref{conj:pq} has been proven for $M=1$, for $p=1$ and any $M$, and for $p=q$ and any $M$.
\end{remark}
\end{theorem}
Finally, here is the main result of this paper, that determines the relation between the von Neumann and the Wehrl entropies.
\begin{theorem}[Wehrl entropy has Gaussian optimizers]\label{thm:entropy}
The minimum Wehrl entropy among all the quantum states with a given von Neumann entropy is achieved by thermal Gaussian states, i.e. for any quantum state $\hat{\rho}$
\begin{equation}\label{eq:entropy}
W\left(\hat{\rho}\right) \ge M\left(\ln\left(g^{-1}\left(\frac{S\left(\hat{\rho}\right)}{M}\right)+1\right)+1\right)\;,
\end{equation}
with $g$ defined in \eqref{eq:defg}.
\begin{proof}
See \autoref{sec:entropy}.
\end{proof}
\end{theorem}

\section{Proof of \autoref{thm:HA}}\label{sec:HA}
Let us first prove that
\begin{equation}\label{eq:claim1}
\int_{\mathbb{C}^M}f\left(\langle\mathbf{z}|\hat{\rho}|\mathbf{z}\rangle\right)\frac{\mathrm{d}^{2M}z}{\pi^M} \le \liminf_{\kappa\to\infty}\frac{\mathrm{Tr}\;f\left(\kappa^M\mathcal{A}_\kappa^{\otimes M}\left(\hat{\rho}\right)\right)}{\kappa^M}\;.
\end{equation}
The proof is based on the following:
\begin{theorem}[Berezin-Lieb inequality \cite{berezin1972covariant}]\label{thm:HA1}
For any trace-class operator $0\le\hat{A}\le\hat{\mathbb{I}}$ and any convex function $f:\left[0,1\right]\to\mathbb{R}$
\begin{equation}
\int_{\mathbb{C}^M}f\left(\langle\mathbf{z}|\hat{A}|\mathbf{z}\rangle\right)\frac{\mathrm{d}^{2M}z}{\pi^M} \le \mathrm{Tr}\;f\left(\hat{A}\right)\;.
\end{equation}
\begin{proof}
Let us diagonalize $\hat{A}$:
\begin{equation}\label{eq:rhodiag}
\hat{A} = \sum_{n\in\mathbb{N}} a_n\;|\psi_n\rangle\langle\psi_n|\;,\quad 0\le a_n\le1\;,\quad\langle\psi_m|\psi_n\rangle=\delta_{mn}\;,\quad\sum_{n\in\mathbb{N}}|\psi_n\rangle\langle\psi_n|=\hat{\mathbb{I}}\;.
\end{equation}
We then have
\begin{equation}
f\left(\langle\mathbf{z}|\hat{A}|\mathbf{z}\rangle\right) = f\left(\sum_{n\in\mathbb{N}}\left|\langle\mathbf{z}|\psi_n\rangle\right|^2a_n\right) \le \sum_{n\in\mathbb{N}}\left|\langle\mathbf{z}|\psi_n\rangle\right|^2 f\left(a_n\right)\;,
\end{equation}
where we have applied Jensen's inequality to $f$ and we have noticed that the completeness relation for the set $\left\{|\psi_n\rangle\right\}_{n\in\mathbb{N}}$ implies for any $\mathbf{z}\in\mathbb{C}^M$
\begin{equation}
\sum_{n\in\mathbb{N}}\left|\langle\mathbf{z}|\psi_n\rangle\right|^2 = \langle\mathbf{z}|\mathbf{z}\rangle = 1\;.
\end{equation}
It follows that
\begin{equation}
\int_{\mathbb{C}^M}f(\langle\mathbf{z}|\hat{A}|\mathbf{z}\rangle)\frac{\mathrm{d}^{2M}z}{\pi^M} \le \sum_{n\in\mathbb{N}}\int_{\mathbb{C}^M}\left|\langle\mathbf{z}|\psi_n\rangle\right|^2 f(a_n)\frac{\mathrm{d}^{2M}z}{\pi^M} = \sum_{n\in\mathbb{N}}f(a_n) = \mathrm{Tr}\,f(\hat{A}),
\end{equation}
where we have used that for the completeness relation \eqref{eq:complz} for any $n\in\mathbb{N}$
\begin{equation}
\int_{\mathbb{C}^M}\left|\langle\mathbf{z}|\psi_n\rangle\right|^2\frac{\mathrm{d}^{2M}z}{\pi^M} = \langle\psi_n|\psi_n\rangle = 1\;.
\end{equation}
\end{proof}
\end{theorem}
From \autoref{lem:Akbound} we have $0\le\kappa^M\mathcal{A}_\kappa^{\otimes M}\left(\hat{\rho}\right)\le\hat{\mathbb{I}}$.
We can then apply \autoref{thm:HA1} to $\hat{A} = \kappa^M\mathcal{A}_\kappa^{\otimes M}\left(\hat{\rho}\right)$ and get
\begin{align}
\mathrm{Tr}\;f\left(\kappa^M\mathcal{A}_\kappa^{\otimes M}\left(\hat{\rho}\right)\right) &\ge \int_{\mathbb{C}^M}f\left(\kappa^M\langle\mathbf{z}|\mathcal{A}_\kappa^{\otimes M}\left(\hat{\rho}\right)|\mathbf{z}\rangle\right)\frac{\mathrm{d}^{2M}z}{\pi^M}\nonumber\\
&= \int_{\mathbb{C}^M}f\left(\langle\mathbf{z}/\sqrt{\kappa}|\hat{\rho}|\mathbf{z}/\sqrt{\kappa}\rangle\right)\frac{\mathrm{d}^{2M}z}{\pi^M}\nonumber\\
&= \kappa^M\int_{\mathbb{C}^M}f\left(\langle\mathbf{z}|\hat{\rho}|\mathbf{z}\rangle\right)\frac{\mathrm{d}^{2M}z}{\pi^M}\;,
\end{align}
where we have used \autoref{lem:Acoh}.
The claim \eqref{eq:claim1} then follows taking the limit $\kappa\to\infty$.

Let us now prove that
\begin{equation}\label{eq:claim2}
\int_{\mathbb{C}^M}f\left(\langle\mathbf{z}|\hat{\rho}|\mathbf{z}\rangle\right)\frac{\mathrm{d}^{2M}z}{\pi^M} \ge \limsup_{\kappa\to\infty}\frac{\mathrm{Tr}\;f\left(\kappa^M\mathcal{A}_\kappa^{\otimes M}\left(\hat{\rho}\right)\right)}{\kappa^M}\;.
\end{equation}
The proof follows from the following:
\begin{theorem}[Berezin-Lieb inequality \cite{berezin1972covariant}]\label{thm:berezin}
For any convex function $f:[0,1]\to\mathbb{R}$ with $f(0)=0$ and any integrable function $\phi:\mathbb{C}^M\to[0,1]$
\begin{equation}
\mathrm{Tr}\;f\left(\int_{\mathbb{C}^M}\phi(\mathbf{z})\;|\mathbf{z}\rangle\langle\mathbf{z}|\;\frac{\mathrm{d}^{2M}z}{\pi^M}\right) \le \int_{\mathbb{C}^M}f(\phi(\mathbf{z}))\;\frac{\mathrm{d}^{2M}z}{\pi^M}\;.
\end{equation}
\begin{proof}
See e.g. \cite{giovannetti2015majorization}, Appendix B.
\end{proof}
\end{theorem}
Let us define the measure-reprepare channel
\begin{align}\label{eq:defM}
\mathcal{M}_\kappa^{\otimes M}\left(\hat{\rho}\right) &= \int_{\mathbb{C}^M}\langle\mathbf{z}|\hat{\rho}|\mathbf{z}\rangle\;|\sqrt{\kappa}\,\mathbf{z}\rangle\langle\sqrt{\kappa}\,\mathbf{z}|\;\frac{\mathrm{d}^{2M}z}{\pi^M} \nonumber\\
&= \int_{\mathbb{C}^M}\frac{\langle\mathbf{z}/\sqrt{\kappa}|\hat{\rho}|\mathbf{z}/\sqrt{\kappa}\rangle}{\kappa^M}\;|\mathbf{z}\rangle\langle\mathbf{z}|\;\frac{\mathrm{d}^{2M}z}{\pi^M}\;.
\end{align}
We can apply \autoref{thm:berezin} to
\begin{equation}
\phi(\mathbf{z}) = \langle\mathbf{z}/\sqrt{\kappa}|\hat{\rho}|\mathbf{z}/\sqrt{\kappa}\rangle\le1\;,\qquad\mathbf{z}\in\mathbb{C}^M\;.
\end{equation}
We have
\begin{equation}
\mathrm{Tr}\;f\left(\kappa^M\mathcal{M}_\kappa^{\otimes M}\left(\hat{\rho}\right)\right) \le \int_{\mathbb{C}^M}f\left(\langle\mathbf{z}/\sqrt{\kappa}|\hat{\rho}|\mathbf{z}/\sqrt{\kappa}\rangle\right)\frac{\mathrm{d}^{2M}z}{\pi^M}\;,
\end{equation}
hence
\begin{equation}\label{eq:limsup}
\int_{\mathbb{C}^M}f\left(\langle\mathbf{z}|\hat{\rho}|\mathbf{z}\rangle\right)\frac{\mathrm{d}^{2M}z}{\pi^M} \ge \limsup_{\kappa\to\infty}\frac{\mathrm{Tr}\;f\left(\kappa^M\mathcal{M}_\kappa^{\otimes M}\left(\hat{\rho}\right)\right)}{\kappa^M}\;.
\end{equation}
We define for any $\mathbf{z}\in\mathbb{C}^M$ the displacement operator \cite{barnett2002methods}
\begin{equation}
\hat{D}(\mathbf{z}) := \exp\left(\sum_{i=1}^M\left(z_i\,\hat{a}_i^\dag - z_i^*\,\hat{a}_i\right)\right)\;,
\end{equation}
that acts on coherent states as
\begin{equation}\label{eq:Dz}
\hat{D}(\mathbf{z})\;|\mathbf{w}\rangle = e^\frac{\mathbf{w}^\dag\mathbf{z}-\mathbf{z}^\dag\mathbf{w}}{2}\;|\mathbf{z}+\mathbf{w}\rangle\;,\qquad \mathbf{z},\,\mathbf{w}\in\mathbb{C}^M\;.
\end{equation}
Let us define the channel
\begin{align}\label{eq:defN}
\mathcal{N}_\kappa^{\otimes M}\left(\hat{\rho}\right) &= \int_{\mathbb{C}^M}\kappa^M e^{-\kappa|z|^2} \hat{D}(\mathbf{z})\;\hat{\rho}\;{\hat{D}(\mathbf{z})}^\dag\;\frac{\mathrm{d}^{2M}z}{\pi^M}\nonumber\\
&= \int_{\mathbb{C}^M}e^{-|z|^2} \hat{D}\left(\frac{\mathbf{z}}{\sqrt{\kappa}}\right)\;\hat{\rho}\;{\hat{D}\left(\frac{\mathbf{z}}{\sqrt{\kappa}}\right)}^\dag\;\frac{\mathrm{d}^{2M}z}{\pi^M}\;.
\end{align}
\begin{lemma}
For any $\kappa\ge1$
\begin{equation}
\mathcal{M}_\kappa^{\otimes M} = \mathcal{A}_\kappa^{\otimes M}\circ\mathcal{N}_\kappa^{\otimes M}\;.
\end{equation}
\begin{proof}
It is sufficient to prove that $\mathcal{M}_\kappa^{\otimes M}\left(\hat{\rho}\right)$ and $\mathcal{A}_\kappa^{\otimes M}\left(\mathcal{N}_\kappa^{\otimes M}\left(\hat{\rho}\right)\right)$ have the same Husimi $Q$ representation for any quantum state $\hat{\rho}$.
Let us fix $\mathbf{w}\in\mathbb{C}^M$.
On one hand we have from \eqref{eq:defM} and \eqref{eq:cohov}
\begin{align}
\langle\mathbf{w}|\mathcal{M}_\kappa^{\otimes M}\left(\hat{\rho}\right)|\mathbf{w}\rangle &= \int_{\mathbb{C}^M}\langle\mathbf{z}|\hat{\rho}|\mathbf{z}\rangle\left|\langle\sqrt{\kappa}\;\mathbf{z}|\mathbf{w}\rangle\right|^2\frac{\mathrm{d}^{2M}z}{\pi^M}\nonumber\\
&= \int_{\mathbb{C}^M}e^{-\left|\mathbf{w}-\sqrt{\kappa}\;\mathbf{z}\right|^2}\;\langle\mathbf{z}|\hat{\rho}|\mathbf{z}\rangle\;\frac{\mathrm{d}^{2M}z}{\pi^M}\;.
\end{align}
On the other hand we have from \autoref{lem:Acoh} and Eqs. \eqref{eq:defN} and \eqref{eq:Dz}
\begin{align}
\langle\mathbf{w}|\mathcal{A}_\kappa^{\otimes M}\left(\mathcal{N}_\kappa^{\otimes M}\left(\hat{\rho}\right)\right)|\mathbf{w}\rangle &= \mathrm{Tr}\left[\mathcal{N}_\kappa^{\otimes M}\left(\hat{\rho}\right)\;\mathcal{A}_\kappa^{\otimes M\dag}\left(|\mathbf{w}\rangle\langle\mathbf{w}|\right)\right]\nonumber\\
&=\frac{\langle\mathbf{w}/\sqrt{\kappa}|\mathcal{N}_\kappa^{\otimes M}\left(\hat{\rho}\right)|\mathbf{w}/\sqrt{\kappa}\rangle}{\kappa^M}\nonumber\\
&= \int_{\mathbb{C}^M}e^{-\kappa\left|\mathbf{z}\right|^2}\langle\mathbf{w}/\sqrt{\kappa}|\hat{D}(\mathbf{z})\;\hat{\rho}\;{\hat{D}(\mathbf{z})}^\dag|\mathbf{w}/\sqrt{\kappa}\rangle\;\frac{\mathrm{d}^{2M}z}{\pi^M}\nonumber\\
&= \int_{\mathbb{C}^M}e^{-\kappa\left|\mathbf{z}\right|^2}\langle\mathbf{w}/\sqrt{\kappa}-\mathbf{z}|\hat{\rho}|\mathbf{w}/\sqrt{\kappa}-\mathbf{z}\rangle\;\frac{\mathrm{d}^{2M}z}{\pi^M}\nonumber\\
&= \int_{\mathbb{C}^M}e^{-\left|\mathbf{w}-\sqrt{\kappa}\;\mathbf{z}\right|^2}\langle\mathbf{z}|\hat{\rho}|\mathbf{z}\rangle\;\frac{\mathrm{d}^{2M}z}{\pi^M}\;.
\end{align}
\end{proof}
\end{lemma}
\begin{lemma}\label{lem:Nk}
For any quantum state $\hat{\rho}$
\begin{equation}
\lim_{\kappa\to\infty}\left\|\mathcal{N}_\kappa^{\otimes M}\left(\hat{\rho}\right)-\hat{\rho}\right\|_1 = 0\;.
\end{equation}
\begin{proof}
We have from \eqref{eq:defN}
\begin{align}
\left\|\mathcal{N}_\kappa^{\otimes M}\left(\hat{\rho}\right)-\hat{\rho}\right\|_1 &= \left\|\int_{\mathbb{C}^M}e^{-|z|^2} \left( \hat{D}\left(\frac{\mathbf{z}}{\sqrt{\kappa}}\right)\;\hat{\rho}\;{\hat{D}\left(\frac{\mathbf{z}}{\sqrt{\kappa}}\right)}^\dag-\hat{\rho}\right) \frac{\mathrm{d}^{2M}z}{\pi^M}\right\|_1\nonumber\\
&\le \int_{\mathbb{C}^M}e^{-|z|^2} \left\| \hat{D}\left(\frac{\mathbf{z}}{\sqrt{\kappa}}\right)\;\hat{\rho}\;{\hat{D}\left(\frac{\mathbf{z}}{\sqrt{\kappa}}\right)}^\dag-\hat{\rho}\right\|_1 \frac{\mathrm{d}^{2M}z}{\pi^M}\;.
\end{align}
The integrands are dominated by
\begin{equation}
\int_{\mathbb{C}^M}2e^{-|z|^2}\frac{\mathrm{d}^{2M}z}{\pi^M} = 2\;,
\end{equation}
and from \autoref{lem:H1}
\begin{equation}
\lim_{\kappa\to\infty}\left\| \hat{D}\left(\frac{\mathbf{z}}{\sqrt{\kappa}}\right)\;\hat{\rho}\;{\hat{D}\left(\frac{\mathbf{z}}{\sqrt{\kappa}}\right)}^\dag-\hat{\rho}\right\|_1 = 0\;.
\end{equation}
The claim then follows from the dominated convergence theorem.
\end{proof}
\end{lemma}
We have from Klein's inequality applied to $\hat{A}=\kappa^M\mathcal{A}_\kappa^{\otimes M}\left(\mathcal{N}_\kappa\left(\hat{\rho}\right)\right)$ and $\hat{B}=\kappa^M\mathcal{A}_\kappa^{\otimes M}\left(\hat{\rho}\right)$ and \eqref{eq:limsup}
\begin{align}
&\int_{\mathbb{C}^M}f\left(\langle\mathbf{z}|\hat{\rho}|\mathbf{z}\rangle\right)\frac{\mathrm{d}^{2M}z}{\pi^M} \ge \limsup_{\kappa\to\infty}\frac{\mathrm{Tr}\;f\left(\kappa^M\mathcal{A}_\kappa^{\otimes M}\left(\mathcal{N}_\kappa^{\otimes M}\left(\hat{\rho}\right)\right)\right)}{\kappa^M}\nonumber\\
&\ge\limsup_{\kappa\to\infty}\left(\frac{\mathrm{Tr}\;f\left(\kappa^M\mathcal{A}_\kappa^{\otimes M}\left(\hat{\rho}\right)\right)}{\kappa^M} - \left\|\mathcal{A}_\kappa^{\otimes M}\left(\mathcal{N}_\kappa^{\otimes M}\left(\hat{\rho}\right)-\hat{\rho}\right)\right\|_1\left\|f'\right\|_\infty\right)\;.
\end{align}
From the contractivity of the trace norm under quantum channels and from \autoref{lem:Nk} we get
\begin{equation}
\limsup_{\kappa\to\infty}\left\|\mathcal{A}_\kappa^{\otimes M}\left(\mathcal{N}_\kappa^{\otimes M}\left(\hat{\rho}\right)-\hat{\rho}\right)\right\|_1 \le \limsup_{\kappa\to\infty}\left\|\mathcal{N}_\kappa^{\otimes M}\left(\hat{\rho}\right)-\hat{\rho}\right\|_1 = 0\;,
\end{equation}
and the claim \eqref{eq:claim2} follows.

\section{Proof of \autoref{thm:maj}}\label{sec:maj}
From \autoref{thm:HA}
\begin{equation}
\int_{\mathbb{C}}f\left(\langle z|\hat{\rho}|z\rangle\right)\frac{\mathrm{d}^{2}z}{\pi} = \lim_{\kappa\to\infty}\frac{\mathrm{Tr}\;f\left(\kappa\;\mathcal{A}_\kappa\left(\hat{\rho}\right)\right)}{\kappa}\;.
\end{equation}
From \autoref{thm:majaux}
\begin{equation}
\mathrm{Tr}\;f\left(\kappa\;\mathcal{A}_\kappa\left(\hat{\rho}\right)\right) \le \mathrm{Tr}\;f\left(\kappa\;\mathcal{A}_\kappa\left(\hat{\rho}^\downarrow\right)\right)\;.
\end{equation}
We then have
\begin{equation}
\int_{\mathbb{C}}f\left(\langle z|\hat{\rho}|z\rangle\right)\frac{\mathrm{d}^{2}z}{\pi} \le \lim_{\kappa\to\infty}\frac{\mathrm{Tr}\;f\left(\kappa\;\mathcal{A}_\kappa\left(\hat{\rho}^\downarrow\right)\right)}{\kappa} = \int_{\mathbb{C}}f\left(\langle z|\hat{\rho}^\downarrow|z\rangle\right)\frac{\mathrm{d}^{2}z}{\pi}\;.
\end{equation}

\section{Proof of \autoref{thm:pq}}\label{sec:pq}
Choosing $f(x)=x^q$ in \autoref{thm:HA} we get
\begin{equation}\label{eq:Qq}
\left\|Q\left(\hat{\rho}\right)\right\|_q^q = \int_{\mathbb{C}^M}{\langle\mathbf{z}|\hat{\rho}|\mathbf{z}\rangle}^q\;\frac{\mathrm{d}^{2M}z}{\pi^M} = \lim_{\kappa\to\infty}\kappa^{M\left(q-1\right)}\mathrm{Tr}\;{\mathcal{A}_\kappa^{\otimes M}\left(\hat{\rho}\right)}^q\;.
\end{equation}
Conjecture \ref{conj:pq} then gives
\begin{equation}
\frac{\left\|Q\left(\hat{\rho}\right)\right\|_q}{\left\|\hat{\rho}\right\|_p} = \lim_{\kappa\to\infty}\kappa^{M\frac{q-1}{q}}\frac{\left\|\mathcal{A}_\kappa^{\otimes M}\left(\hat{\rho}\right)\right\|_q}{\left\|\hat{\rho}\right\|_p} \le \left(\lim_{\kappa\to\infty}\sup_{0\le z<1}\kappa^{\frac{q-1}{q}}\frac{\left\|\mathcal{A}_\kappa\left(\hat{\omega}_z\right)\right\|_q}{\left\|\hat{\omega}_z\right\|_p}\right)^M\;.
\end{equation}
We can compute from \eqref{eq:omegaz} for any $0\le z<1$ and any $p\ge1$
\begin{equation}\label{eq:omegap}
\left\|\hat{\omega}_z\right\|_p = \frac{1-z}{\left(1-z^p\right)^\frac{1}{p}}\;.
\end{equation}
We have from \eqref{eq:Az} and \eqref{eq:omegap} for any $0\le z<1$ and any $\kappa\ge1$
\begin{equation}\label{eq:norm}
\kappa^{\frac{q-1}{q}}\left\|\mathcal{A}_\kappa\left(\hat{\omega}_z\right)\right\|_q = \frac{1-z}{\left(\kappa-\kappa\left(1-\frac{1-z}{\kappa}\right)^q\right)^\frac{1}{q}}\;.
\end{equation}
From \eqref{eq:Qq} with $\hat{\rho}=\hat{\omega}_z$ we get
\begin{equation}\label{eq:norm2}
\left\|Q\left(\hat{\omega}_z\right)\right\|_q = \lim_{\kappa\to\infty}\kappa^{\frac{q-1}{q}}\left\|\mathcal{A}_\kappa\left(\hat{\omega}_z\right)\right\|_q = \frac{\left(1-z\right)^\frac{q-1}{q}}{q^\frac{1}{q}}\;.
\end{equation}
Let us choose $1\le r<q$.
For any $0\le z<1$ and any $\kappa\ge\frac{1}{x_r}$, with $x_r$ as in \autoref{lem:q}, we have
\begin{equation}
\frac{1-z}{\kappa}\le\frac{1}{\kappa}\le x_r\;.
\end{equation}
We then have from \autoref{lem:q}
\begin{equation}
\left(1-\frac{1-z}{\kappa}\right)^q \le 1-r\frac{1-z}{\kappa}\;,
\end{equation}
and from \eqref{eq:norm} and \eqref{eq:norm2}
\begin{equation}
\kappa^{\frac{q-1}{q}}\left\|\mathcal{A}_\kappa\left(\hat{\omega}_z\right)\right\|_q \le \frac{\left(1-z\right)^\frac{q-1}{q}}{r^\frac{1}{q}} = \frac{q^\frac{1}{q}}{r^\frac{1}{q}}\left\|Q\left(\hat{\omega}_z\right)\right\|_q\;.
\end{equation}
It follows that
\begin{equation}
\lim_{\kappa\to\infty}\sup_{0\le z<1}\kappa^{\frac{q-1}{q}}\frac{\left\|\mathcal{A}_\kappa\left(\hat{\omega}_z\right)\right\|_q}{\left\|\hat{\omega}_z\right\|_p} \le \frac{q^\frac{1}{q}}{r^\frac{1}{q}}\sup_{0\le z<1}\frac{\left\|Q\left(\hat{\omega}_z\right)\right\|_q}{\left\|\hat{\omega}_z\right\|_p}\;,
\end{equation}
and the claim follows taking the limit $r\to q$.

Let us prove that the supremum in \eqref{eq:pq} is finite if $1\le p \le q$.
We have for any $0\le z <1$
\begin{equation}
\frac{\left(1-z^p\right)^\frac{1}{p}}{\left(1-z\right)^\frac{1}{q}} \le \left(\frac{1-z^p}{1-z}\right)^\frac{1}{p} \le p^\frac{1}{p}\;.
\end{equation}
On the other hand, if $p>q\ge1$ we have
\begin{equation}
\lim_{z\to 1}\frac{\left(1-z^p\right)^\frac{1}{p}}{\left(1-z\right)^\frac{1}{q}} = \infty\;,
\end{equation}
and the supremum in \eqref{eq:pq} is infinite.

\section{Proof of \autoref{thm:entropy}}\label{sec:entropy}
Let us show that \eqref{eq:entropy} is saturated by the thermal Gaussian states $\hat{\omega}_z^{\otimes M}$, $0\le z<1$.
On one hand we have from \eqref{eq:defg} $S\left(\hat{\omega}_z^{\otimes M}\right) = M\,g(E)$, with $E$ given by \eqref{eq:Ez}.
On the other hand we have
\begin{equation}
Q\left(\hat{\omega}_z^{\otimes M}\right)(\mathbf{z}) = \frac{e^{-\frac{|\mathbf{z}|^2}{E+1}}}{\left(E+1\right)^M}\;,
\end{equation}
and
\begin{equation}\label{eq:SQG}
W\left(\hat{\omega}_z^{\otimes M}\right) = M\left(\ln\left(E+1\right)+1\right)\;.
\end{equation}

We will prove \autoref{thm:entropy} by induction on $M$.
Let us prove the claim for $M=1$.
The proof of \eqref{eq:claim1} does not require $f$ to be differentiable.
We can then choose $f(x)=x\ln x$ and get
\begin{equation}
W\left(\hat{\rho}\right) \ge \limsup_{\kappa\to\infty}\left(S\left(\mathcal{A}_\kappa\left(\hat{\rho}\right)\right)-\ln\kappa\right)\;.
\end{equation}
We get from \autoref{thm:epni}
\begin{align}
W\left(\hat{\rho}\right) &\ge \limsup_{\kappa\to\infty}\left(g\left(\kappa\;g^{-1}\left(S\left(\hat{\rho}\right)\right)+\kappa-1\right)-\ln\kappa\right)\nonumber\\
&= \ln\left(g^{-1}\left(S\left(\hat{\rho}\right)\right)+1\right)+1\;,
\end{align}
where we have used that for $x\to\infty$
\begin{equation}
g(x) = \ln x + 1 + \mathcal{O}\left(\frac{1}{x}\right)\;.
\end{equation}

From the inductive hypothesis, we can assume that \autoref{thm:entropy} holds for a given $M$.
It is then sufficient to prove the claim for $M+1$.
Let $\hat{\rho}$ be a $(M+1)$-mode quantum state.
We have from the chain rule for the Shannon differential entropy
\begin{align}
&W\left(\hat{\rho}\right) = W\left(\mathrm{Tr}_{M+1}\hat{\rho}\right)\nonumber\\
&+ \int_{\mathbb{C}^M}W\left(\frac{\langle z_1\ldots z_M|\hat{\rho}|z_1\ldots z_M\rangle}{\mathrm{Tr}\langle z_1\ldots z_M|\hat{\rho}|z_1\ldots z_M\rangle}\right)\mathrm{Tr}\langle z_1\ldots z_M|\hat{\rho}|z_1\ldots z_M\rangle\,\frac{\mathrm{d}z_1\ldots\mathrm{d}z_M}{\pi^M}\;,
\end{align}
where $\mathrm{Tr}_{M+1}\hat{\rho}$ is the $M$-mode quantum state given by the partial trace of $\hat{\rho}$ over the last mode.
We have from the inductive hypothesis
\begin{equation}
W\left(\mathrm{Tr}_{M+1}\hat{\rho}\right) \ge M f\left(\frac{S\left(\mathrm{Tr}_{M+1}\hat{\rho}\right)}{M}\right)\;,
\end{equation}
where we have defined for any $x\ge0$
\begin{equation}\label{eq:deff}
f(x) := \ln\left(g^{-1}(x) + 1\right) + 1\;.
\end{equation}
\autoref{thm:entropy} for $M=1$ implies for any $z_1,\,\ldots,\,z_M\in\mathbb{C}$
\begin{equation}
W\left(\frac{\langle z_1\ldots z_M|\hat{\rho}|z_1\ldots z_M\rangle}{\mathrm{Tr}\langle z_1\ldots z_M|\hat{\rho}|z_1\ldots z_M\rangle}\right) \ge f\left(S\left(\frac{\langle z_1\ldots z_M|\hat{\rho}|z_1\ldots z_M\rangle}{\mathrm{Tr}\langle z_1\ldots z_M|\hat{\rho}|z_1\ldots z_M\rangle}\right)\right)\;.
\end{equation}
We then have
\begin{align}\label{eq:chain}
&W\left(\hat{\rho}\right) \ge M f\left(\frac{S\left(\mathrm{Tr}_{M+1}\hat{\rho}\right)}{M}\right)\nonumber\\
&+\int_{\mathbb{C}^M}f\left(S\left(\frac{\langle z_1\ldots z_M|\hat{\rho}|z_1\ldots z_M\rangle}{\mathrm{Tr}\langle z_1\ldots z_M|\hat{\rho}|z_1\ldots z_M\rangle}\right)\right)\mathrm{Tr}\langle z_1\ldots z_M|\hat{\rho}|z_1\ldots z_M\rangle\,\frac{\mathrm{d}z_1\ldots\mathrm{d}z_M}{\pi^M}\nonumber\\
&\ge M f\left(\frac{S\left(\mathrm{Tr}_{M+1}\hat{\rho}\right)}{M}\right)\nonumber\\
&+ f\left(\int_{\mathbb{C}^M}S\left(\frac{\langle z_1\ldots z_M|\hat{\rho}|z_1\ldots z_M\rangle}{\mathrm{Tr}\langle z_1\ldots z_M|\hat{\rho}|z_1\ldots z_M\rangle}\right)\mathrm{Tr}\langle z_1\ldots z_M|\hat{\rho}|z_1\ldots z_M\rangle\,\frac{\mathrm{d}z_1\ldots\mathrm{d}z_M}{\pi^M}\right)\;,
\end{align}
where we have used that $f$ is convex (see \autoref{lem:conv}).
The argument of $f$ in the last line of \eqref{eq:chain} is the entropy of the last mode of $\hat{\rho}$ conditioned on the outcomes of the heterodyne measurements on the first $M$ modes of $\hat{\rho}$.
We then have from the data-processing inequality for the quantum conditional entropy applied to the heterodyne measurement of the first $M$ modes of $\hat{\rho}$
\begin{align}
&\int_{\mathbb{C}^M}S\left(\frac{\langle z_1\ldots z_M|\hat{\rho}|z_1\ldots z_M\rangle}{\mathrm{Tr}\langle z_1\ldots z_M|\hat{\rho}|z_1\ldots z_M\rangle}\right)\mathrm{Tr}\langle z_1\ldots z_M|\hat{\rho}|z_1\ldots z_M\rangle\,\frac{\mathrm{d}z_1\ldots\mathrm{d}z_M}{\pi^M}\nonumber\\
&\ge S\left(\hat{\rho}\right) - S\left(\mathrm{Tr}_{M+1}\hat{\rho}\right)\;.
\end{align}
Finally, since $f$ is increasing we have
\begin{equation}
W\left(\hat{\rho}\right) \ge M f\left(\frac{S\left(\mathrm{Tr}_{M+1}\hat{\rho}\right)}{M}\right) + f\left(S\left(\hat{\rho}\right) - S\left(\mathrm{Tr}_{M+1}\hat{\rho}\right)\right) \ge \left(M+1\right)f\left(\frac{S\left(\hat{\rho}\right)}{M+1}\right)\;,
\end{equation}
where we have used the convexity of $f$ again.

\section{Conclusions}\label{sec:concl}
We have proven that the quantum-classical channel that associates to a quantum state its Husimi $Q$ representation is asymptotically equivalent to the Gaussian quantum-limited amplifier with infinite amplification parameter (\autoref{thm:HA}).
This equivalence has permitted us to determine the minimum Wehrl entropy among all the quantum states with a given von Neumann entropy, and prove that it is achieved by a thermal Gaussian state (\autoref{thm:entropy}).
This result determines the relation between the von Neumann and the Wehrl entropies.
The same equivalence has also permitted us to determine the $p\to q$ norms of the aforementioned quantum-classical channel in the two particular cases of one mode and $p=q$, and prove that they are achieved by thermal Gaussian states (\autoref{thm:pq}).
A proof of Conjecture \ref{conj:pq} for any $M$, $p$ and $q$ would determine the $p\to q$ norms of this quantum-classical channel for any $M$, $p$ and $q$.

The Husimi $Q$ representation of a quantum state coincides with the probability distribution of the outcome of a heterodyne measurement performed on the state.
Then, our results can find applications in quantum cryptography for the quantum key distribution schemes based on the heterodyne measurement \cite{weedbrook2012gaussian,weedbrook2004quantum}.

\begin{acknowledgements}
I thank Jan Philip Solovej for very fruitful discussions and for a careful reading of this paper.
\end{acknowledgements}

\appendix
\section{The Gaussian quantum-limited attenuator}\label{app:E}
The $M$-mode Gaussian quantum-limited attenuator $\mathcal{E}_{\lambda}^{\otimes M}$ with attenuation parameter $0\le\lambda\le1$ is the quantum channel that mixes through a beamsplitter with transmissivity $\lambda$ the input state $\hat{\rho}$ and the vacuum state of a $M$-mode ancillary Gaussian system $B$ with ladder operators $\hat{b}_1,\ldots,\,\hat{b}_M$:
\begin{equation}
\mathcal{E}_{\lambda}^{\otimes M}\left(\hat{\rho}\right)= \mathrm{Tr}_B\left[\hat{U}_\lambda\left(\hat{\rho}\otimes|\mathbf{0}\rangle\langle\mathbf{0}|\right)\hat{U}_\lambda^\dag\right]\;.
\end{equation}
The beamsplitter is implemented by the two-mode mixing unitary operator
\begin{equation}
\hat{U}_\lambda=\exp\left(\arccos\sqrt{\lambda}\sum_{i=1}^M\left(\hat{a}_i^\dag\,\hat{b}_i-\hat{a}_i\,\hat{b}_i^\dag\right)\right)\;,
\end{equation}
and it acts on the ladder operators as
\begin{align}
\hat{U}_\lambda^\dag\;\hat{a}_i\;\hat{U}_\lambda &= \sqrt{\lambda}\;\hat{a}_i + \sqrt{1-\lambda}\;\hat{b}_i\;,\\
\hat{U}_\lambda^\dag\;\hat{b}_i\;\hat{U}_\lambda &= -\sqrt{1-\lambda}\;\hat{a}_i + \sqrt{\lambda}\;\hat{b}_i\;,\qquad i=1,\ldots,\,M\;.
\end{align}

\begin{lemma}[\cite{holevo2015gaussian}]\label{lem:Ecoh}
The Gaussian quantum-limited attenuator preserves the set of coherent states, i.e. for any $0\le\lambda\le 1$ and any $\mathbf{z}\in\mathbb{C}^M$
\begin{equation}
\mathcal{E}_\lambda^{\otimes M}\left(|\mathbf{z}\rangle\langle\mathbf{z}|\right) = |\sqrt{\lambda}\,\mathbf{z}\rangle\langle\sqrt{\lambda}\,\mathbf{z}|\;.
\end{equation}
\end{lemma}
The relation with the amplifier is given by
\begin{theorem}[\cite{ivan2011operator}, Theorem 9]\label{thm:dual}
The Gaussian quantum-limited attenuator and amplifier are mutually dual, i.e. for any $\kappa\ge1$
\begin{equation}
\kappa^M\mathcal{A}_\kappa^{\otimes M\dag} = \mathcal{E}^{\otimes M}_\frac{1}{\kappa}\;.
\end{equation}
\end{theorem}
\begin{lemma}\label{lem:Acoh}
For any $\kappa\ge1$ and any $\mathbf{z}\in\mathbb{C}^M$
\begin{equation}
\kappa^M\mathcal{A}_\kappa^{\otimes M\dag}\left(|\mathbf{z}\rangle\langle\mathbf{z}|\right) = |\mathbf{z}/\sqrt{\kappa}\rangle\langle\mathbf{z}/\sqrt{\kappa}|\;.
\end{equation}
\begin{proof}
Follows from \autoref{thm:dual} and \autoref{lem:Ecoh}.
\end{proof}
\end{lemma}
\begin{lemma}\label{lem:Akbound}
For any quantum state $\hat{\rho}$ and any $\kappa\ge1$
\begin{equation}
0\le \kappa^M\mathcal{A}_\kappa^{\otimes M}\left(\hat{\rho}\right)\le\hat{\mathbb{I}}\;.
\end{equation}
\begin{proof}
We have from \autoref{thm:dual}
\begin{equation}
0\le\kappa^M\mathcal{A}_\kappa^{\otimes M}\left(\hat{\rho}\right) = \mathcal{E}_\frac{1}{\kappa}^{\otimes M\dag}\left(\hat{\rho}\right) \le \mathcal{E}_\frac{1}{\kappa}^{\otimes M\dag}\left(\hat{\mathbb{I}}\right) = \hat{\mathbb{I}}\;,
\end{equation}
where we have used that, since the Gaussian quantum-limited attenuator is trace-preserving, its dual is unital.
\end{proof}
\end{lemma}

\section{Auxiliary theorems and lemmas}\label{app:aux}
\begin{theorem}[Klein's inequality]
Let $f\in C^1([0,1])$ be a real convex function with $f(0)=0$.
Then, for any two trace-class operators $0\le\hat{A},\hat{B}\le\hat{\mathbb{I}}$
\begin{equation}
\mathrm{Tr}\;f\left(\hat{B}\right) \le \mathrm{Tr}\;f\left(\hat{A}\right) + \left\|\hat{B}-\hat{A}\right\|_1\left\|f'\right\|_\infty\;.
\end{equation}
\begin{proof}
Let us diagonalize $\hat{A}$ and $\hat{B}$:
\begin{align}
\hat{A}&=\sum_{m\in\mathbb{N}}a_m|\phi_m\rangle\langle\phi_m|,\quad0\le a_m\le1,\quad \langle\phi_m|\phi_{m'}\rangle = \delta_{mm'},\quad\sum_{m\in\mathbb{N}}|\phi_m\rangle\langle\phi_m|=\hat{\mathbb{I}}\nonumber\\
\hat{B}&=\sum_{n\in\mathbb{N}}b_n|\psi_n\rangle\langle\psi_n|,\quad0\le b_n\le1,\quad \langle\psi_n|\psi_{n'}\rangle = \delta_{nn'},\quad\sum_{n\in\mathbb{N}}|\psi_n\rangle\langle\psi_n|=\hat{\mathbb{I}}\;.
\end{align}
Since $f$ is convex, for any $0\le a,b\le1$
\begin{equation}
f(b) \le f(a) + \left(b-a\right)f'(b)\;.
\end{equation}
We then have
\begin{align}
\mathrm{Tr}\;f\left(\hat{B}\right) &= \sum_{n\in\mathbb{N}}f(b_n) = \sum_{m,n\in\mathbb{N}}\left|\langle\phi_m|\psi_n\rangle\right|^2f(b_n)\nonumber\\
&\le \sum_{m,n\in\mathbb{N}}\left|\langle\phi_m|\psi_n\rangle\right|^2\left(f(a_m) + \left(b_n-a_m\right)f'(b_n)\right)\nonumber\\
&= \sum_{m\in\mathbb{N}}f(a_m) + \sum_{n\in\mathbb{N}}b_n\,f'(b_n) - \sum_{m,n\in\mathbb{N}}\left|\langle\phi_m|\psi_n\rangle\right|^2 a_m\,f'(b_n)\nonumber\\
&=\mathrm{Tr}\left[f\left(\hat{A}\right)+\left(\hat{B}-\hat{A}\right)f'\left(\hat{B}\right)\right]\nonumber\\
&\le \mathrm{Tr}\;f\left(\hat{A}\right) + \left\|\hat{B}-\hat{A}\right\|_1\left\|f'\right\|_\infty\;.
\end{align}
\end{proof}
\end{theorem}

\begin{lemma}\label{lem:H1}
For any quantum state $\hat{\rho}$
\begin{equation}
\lim_{\mathbf{z}\to\mathbf{0}}\left\| \hat{D}(\mathbf{z})\;\hat{\rho}\;{\hat{D}(\mathbf{z})}^\dag-\hat{\rho}\right\|_1 = 0\;.
\end{equation}
\begin{proof}
Let us diagonalize $\hat{\rho}$:
\begin{equation}
\hat{\rho} = \sum_{n\in\mathbb{N}} p_n\;|\psi_n\rangle\langle\psi_n|\;,\quad p_n\ge0\;,\quad\sum_{n\in\mathbb{N}}p_n=1\;,\quad \langle\psi_m|\psi_n\rangle=\delta_{mn}\;.
\end{equation}
We have for any $\mathbf{z}\in\mathbb{C}^M$
\begin{align}
\left\| \hat{D}(\mathbf{z})\;\hat{\rho}\;{\hat{D}(\mathbf{z})}^\dag-\hat{\rho}\right\|_1 &= \left\|\sum_{n\in\mathbb{N}}p_n\left(\hat{D}(\mathbf{z})|\psi_n\rangle\langle\psi_n|{\hat{D}(\mathbf{z})}^\dag - |\psi_n\rangle\langle\psi_n|\right)\right\|_1\nonumber\\
&\le \sum_{n\in\mathbb{N}}p_n\left\|\hat{D}(\mathbf{z})|\psi_n\rangle\langle\psi_n|{\hat{D}(\mathbf{z})}^\dag - |\psi_n\rangle\langle\psi_n|\right\|_1\nonumber\\
&=\sum_{n\in\mathbb{N}}2p_n\sqrt{1-\left|\langle\psi_n|\hat{D}(\mathbf{z})|\psi_n\rangle\right|^2}\;.
\end{align}
The sums are dominated by
\begin{equation}
\sum_{n\in\mathbb{N}}2p_n = 2\;.
\end{equation}
Since $\hat{D}(\mathbf{z})$ is strongly continuous in $\mathbf{z}$ \cite{holevo2013quantum}, it is also weakly continuous, and we have for any $n\in\mathbb{N}$
\begin{equation}
\lim_{\mathbf{z}\to\mathbf{0}}\langle\psi_n|\hat{D}(\mathbf{z})|\psi_n\rangle = 1\;.
\end{equation}
The claim then follows from the dominated convergence theorem.
\end{proof}
\end{lemma}
\begin{lemma}\label{lem:q}
For any $1\le r<q$ and any $0\le x\le x_r$, with
\begin{equation}
x_r := 1-\left(\frac{r}{q}\right)^\frac{1}{q-1}>0\;,
\end{equation}
we have
\begin{equation}
\left(1-x\right)^q \le 1-r\,x\;.
\end{equation}
\begin{proof}
Let us define
\begin{equation}
\phi(x):=1-r\,x -\left(1-x\right)^q\;.
\end{equation}
We have $\phi(0)=0$, and
\begin{equation}
\phi'(x) = q\left(1-x\right)^{q-1}-r\;.
\end{equation}
The claim follows since $\phi'(x)\ge0$ for any $0\le x\le x_r$.
\end{proof}
\end{lemma}
\begin{equation}
\left\|\hat{\omega}_z\right\|_p = \frac{1-z}{\left(1-z^p\right)^\frac{1}{p}}
\end{equation}
\begin{equation}
w = 1 - \frac{1-z}{\kappa}\;,\qquad 1-w = \frac{1-z}{\kappa}
\end{equation}

\begin{lemma}\label{lem:conv}
The function $f$ defined in \eqref{eq:deff} is increasing and convex.
\begin{proof}
Since $g$ is increasing, also $g^{-1}$ is increasing, and $f$ is increasing.
We will prove that $f''(x)\ge0$ for any $x>0$.
We have for any $y>0$
\begin{equation}
f''(g(y)) = -\frac{g''(y)}{{g'(y)}^3}\,\frac{1-h(y)}{1+y}\;,
\end{equation}
where
\begin{equation}
h(y) := -\frac{g'(y)}{\left(1+y\right)g''(y)} = y\ln\left(1+\frac{1}{y}\right) \le 1\;.
\end{equation}
The claim then follows since $g$ is concave and $g''(y)<0$.
\end{proof}
\end{lemma}

\bibliographystyle{spmpsci}
\bibliography{biblio}
\end{document}